# Bernard Yarnton Mills 1920-2011[1]

*R. H. Frater,*[A] *W. M. Goss*[B] *and H. W. Wendt*[C]

(To appear in 'Historical Records of Australian Science', Vol. 24, No. 2, December 2013)

[A] Corresponding author. P.O. Box 456, Lindfield, NSW 2070, Australia. E-mail: Bob.Frater@resmed.com.au.
[B] National Radio Astronomy Observatory, PO Box 0, Socorro, New Mexico 87801, USA.
[C] 18 Boambillee Avenue, Vaucluse NSW 2030, Australia.

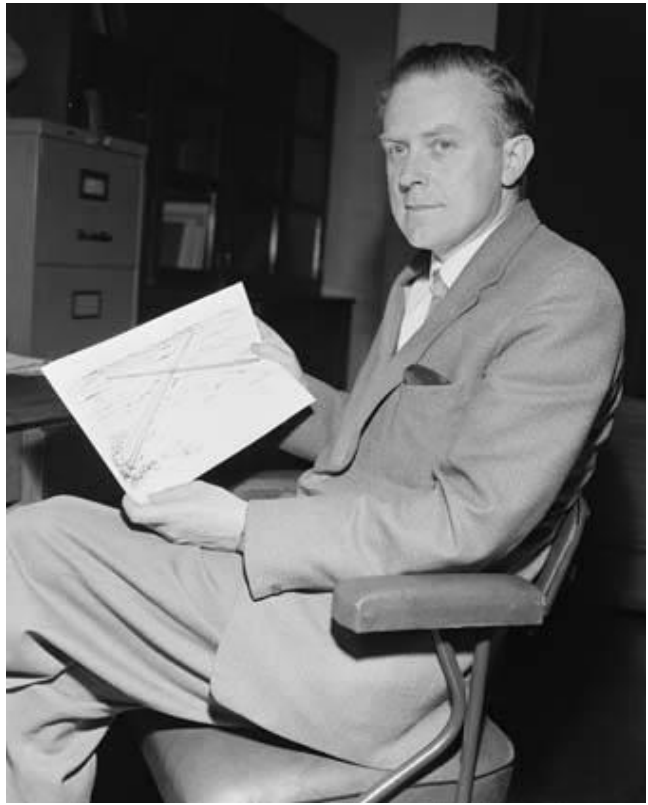

Bernie Mills is remembered globally as an influential pioneer in the evolving field of radio astronomy. His contributions with the 'Mills Cross' at the CSIRO Division of Radiophysics and later at the University of Sydney's School of Physics and the development of the Molonglo Observatory Synthesis Telescope (MOST) were widely recognised as astronomy evolved in the years 1948-1985 and radio astronomy changed the viewpoint of the astronomer as a host of new objects were discovered.

**Introduction: the Legacy of Bernard Mills**

Bernie Mills stands as a giant of twentieth-century radio astronomy. In both the Australian and international contexts, he is remembered as an influential pioneer in the early evolution of this field. His contributions at Australia's Commonwealth Scientific and Industrial Research Organisation (CSIRO), in its Division of Radiophysics, and later at University of Sydney's School of Physics were widely recognised as astronomy evolved in the years 1948-85. His desire for physical understanding in all he did underpinned his achievements. Radio astronomy, the second window on the Universe, changed the viewpoint of the astronomer as a host of new objects were discovered. In this period the eruptive solar corona, radio galaxies, the structure of the Milky Way, quasars, black holes, pulsars, dark matter and the relic radiation from the big bang were discovered.

---

[1] A similar memoir will appear in *Biographical Memoirs of Fellows of the Royal Society of London*, vol. 59, 2013.

Bernie's career can be considered in two broad phases. The first deals with his time at the CSIRO and its predecessor, the Council for Scientific and Industrial Research (CSIR), where, a newly qualified engineer, he became an astronomer. During this period he made a fundamental contribution to the elucidation of the nature of the newly-discovered discrete astronomical radio sources, as well inventing the cross-type radio telescope that influenced developments not only in radio astronomy but also in other fields such as sonar and radar. In the second stage of his career, while at University of Sydney, he returned to the role of the engineer, concentrating on the construction of the Molonglo Cross and later the Molonglo Observatory Synthesis Telescope (MOST) which remains in operation to this date. In his retirement years Bernie remained in active contact with many colleagues in Australia and overseas. His autobiographical text of 2006 (92) remains a striking monument to a career characterized by innovation and insight.

**Early Life**

Bernard Yarnton Mills was born at the Sydney seaside suburb of Manly on 8 August 1920, the only child of Ellice Yarnton Mills (1882-1945) and Sylphide Mills née Dinwiddie (1890-1962).

Ellice Mills was an architect employed by the Sydney Municipal Council who had migrated to Australia from England shortly before the First World War. Bernie described the family background as 'church and also business'. His grandfather, Daniel Yarnton Mills, had been an actuary and a very strong and well-known chess player, eight times Scottish chess master and British Amateur Champion in 1890 as well as being undefeated while representing Great Britain in all the American-English cable matches between 1896 and 1911. Bernie described his father as a successful golfer and watercolour artist.

Sylphide Dinwiddie was from New Zealand and had come to Australia as a teacher of dance. On a return visit to New Zealand she was introduced to Christian Science, subsequently bringing up Bernie as a Christian Scientist. This had a profound effect on his development.

A precocious child, Bernie was sent to the Kings School, Parramatta, a school in the English tradition in Sydney. He found his school years unrewarding. He was ahead of his age group, skipped a class and again headed his class which proved difficult socially in a school devoted to sporting ability rather than scholastic prowess. The one bright spot in this otherwise difficult period of his life was the physics master, 'Edgy' Bartlett, who was an enthusiastic educator and communicator. Mathematics teaching was uninspiring and Bernie regarded his grounding in the subject as poor, but this encouraged the development of the physical understanding that shone throughout Bernie's career.

**University of Sydney**

Bernie's father wanted him to go to the Duntroon Military College but eventually relented and allowed him to take up a scholarship to do engineering at the University of Sydney. He began his university studies in 1937 at the age of 16, initially focusing on physics and later moving to engineering.

Bernie warmed to the university environment where he met people from very different social and political backgrounds and where his intellectual capabilities were challenged.  He followed the lead of his grandfather into a serious interest in chess – somewhat at the expense of his studies – winning University and City of Sydney championships.  During university vacations he undertook work experience with several engineering firms where he gained practical experience and exposure to a broad range of workers and left-wing politics far removed from the environment of his upbringing.

In 1939, Bernie met his future wife, Russian-born Lerida Karmalsky, a medical student and chess player who introduced him to the Labor Club. They were married in 1942. Like many left-wing students of the time, they both joined the Communist Party  but they eventually resigned in disgust after the Russian invasion of Hungary in 1956.  (By this time, Bernie had been with the CSIRO for thirteen years.)

Bernie was awarded a Bachelor of Science degree in Mathematics and Physics in 1941 and a Bachelor of Engineering degree with second-class honours in 1943.  He was proud of the fact that he and Ron Bracewell, a fellow student who would also later become a distinguished radio astronomer, had gone on to great career success with Second Class Honours degrees.

**The Division of Radiophysics**

On completing their studies at the end of 1942, all six honours students of his graduation group were recruited into CSIR's Division of Radiophysicsto work on radar research and development.  Bernie worked in the receiver and display group where he contributed to the design of a height-finding radar that was used after the war at Mascot airport in Sydney.  The rapidly rotating helical scan system was influenced by work he did with Ruby Payne-Scott (who would go on to play a key role in early solar radio astronomy [see Goss and McGee, 2009, Goss, 2013)]) on faint plan-position indicator (PPI) radar signals, concepts that proved very useful in later years.  The development of this radar 'was perhaps RPL's [the Radiophysics Laboratory's] outstanding technical achievement during the war' (Minnett, 1999, p. 465).

Joseph ('Joe') Pawsey played a very important role in Bernie's development.  Pawsey headed the general development and experimental work at RPL at this time and would go on to become the highly-regarded leader of the radio astronomy group.  In the short lecture courses that he gave on transmission lines and antennas, he promoted a physical understanding rather than the highly mathematical approach to which Bernie had been exposed during his studies.  This physical understanding was a key in Bernie's career and life.

## After the War

A number of different research activities were undertaken by the Radiophysics Division after the war. Edward 'Taffy' Bowen, who had become Chief of the Division, had the idea of using a magnetron as a power source for a linear electron accelerator (Bowen et al., 1946). Although the development of linear accelerators was soon abandoned, Bernie was given the task of exploring the possibilities of developing the equipment for X-ray work (Home, 2006). This was his first exposure to physics that went beyond Maxwell's equations and proved a practical introduction to relativity and even some quantum physics.

The development was successful and resulted in Bernie's first major publication, 'A Million Volt Resonant Cavity X-ray Tube' (5). While his project report was submitted as a Master of Engineering thesis and awarded first-class honours in 1950, it was decided that there was no future in Australia for this kind of research and Bernie was moved into a group developing a digital computer. However, in late 1947 he was diagnosed with early-stage tuberculosis. Fortunately this potentially life-threatening condition was detected early and after six months' bed rest he made a full recovery.

## Engineer becomes Astronomer

The timing of his return to work gave Bernie the choice of continuing with the computer development or joining Pawsey's new radio astronomy group. He jumped at the opportunity to work again with Pawsey.

Bernie describes his entry into the group: 'I began with some necessary study of astronomy, some equipment development, and assistance to Chris Christiansen and Don Yabsley in their observations of the 1948 Solar eclipse [Christiansen et al., 1949a, Christiansen et al., 1949b]). This led to my first paper in radio astronomy, as the junior author! However, I had to get started on a project of my own and Pawsey suggested two possibilities, to try and detect the H-line which had been predicted by van de Hulst or to use the swept-lobe solar interferometer, being developed by Alec Little and Ruby Payne-Scott [1951]), to study the mysterious radio sources which John Bolton had made his own using a cliff-top interferometer[2] Pawsey was very dubious about the future of this form of interferometry and felt that the two-antenna interferometer of Payne-Scott and Little offered much the better prospects for accurate positions and identifications even though the antennas were smaller (simple Yagis). If I had been a trained astronomer, no doubt I would have chosen the H-line project but I was intrigued by the mystery surrounding the 'discrete sources' and had no hesitation in making this choice'.[3]

---

[2] The sea-cliff interferometer had first been used by Lindsey McCready, Joe Pawsey and Ruby Payne Scott (1947) to investigate the source of solar radio emission. This was in fact the first use of interferometry in radio astronomy. Bolton and his team at Dover Heights went on to use the sea-cliff interferometer for their pioneering work in investigating cosmic radio sources and this resulted in the first three optical identifications of discrete radio sources, two extragalactic and one the Crab Nebula, the remnant of the supernova of 1054 (Bolton et al., 1949).

[3] Unsourced quotations are from Bernie Mills' autobiographical notes, held by the University of Sydney Archives.

Over a period of eight months from May to December 1949, Bernie, with the young English engineer Aidan Thomas, used the 97 MHz swept-lobe interferometer (see Figure 1) located at the Division's Potts Hill field station to determine the astronomical position of the strong discrete radio source Cygnus A (7): 'Eventually we determined a position with probable errors of some two minutes of arc. A photograph of the general area of the Cygnus source had been sent to Bolton by Rudolph Minkowski and on the photograph, after much trouble, we found a small nebulous object within our positional uncertainties. We were confident that this was the source and that it was a Galactic nebulosity because of its location so close to the Galactic plane. However, before publishing I wrote to Minkowski to seek his interpretation, naturally expecting that he would confirm our identification'.

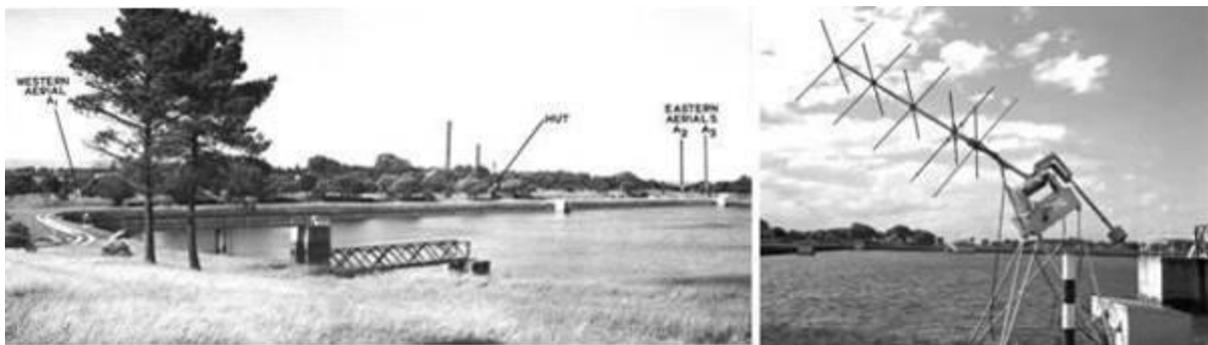

Figure 1: Potts Hill field station and the swept-lobe interferometer used by Mills for observations of Cygnus A. The left figure shows the location of the three aerials of the swept-lobe interferometer and the associated receiver hut. The view is looking to the northeast. The right figure shows a close up view of the western Yagi aerial (Courtesy of the CSIRO Radio Astronomy Image Archive: 2312 & 2217 Image Date: 28 July 1950).

To Bernie's great disappointment, Minkowski did not accept their identification. Although the Cygnus A source rose only some 16 degrees above the horizon, causing problems with ground reflections and likely second-order refraction in the troposphere, they had managed to get their positional error down to 1.1 arc min in right ascension and 3 arc min in declination. Minkowski suggested that this was still too large to provide a definitive identification. Subsequently, Smith (1951) provided an improved position estimate with an error of 0.2 arc min of right ascension and 1 arc min of declination, and with this Baade and Minkowski (1954) confirmed the identification with the faint nebula originally proposed by Mills and Thomas.[4] The whole episode marked the beginning of Bernie's development of a 'healthy scepticism toward authoritative pronouncements and the confidence to rely on my own judgment' (92).

**Badgerys Creek**

The limited sensitivity of the swept-lobe interferometer made it suitable only for observing the strongest discrete sources. Spurred on by his curiosity, Bernie decided to build a new interferometer and continue his investigations. The increasing levels of radio interference at Potts Hill (79) and the need for longer baselines required a new site, which he found at

---

[4] For a full discussion of the identification of Cygnus A, see Sullivan (2009), p. 335.

Badgerys Creek, New South Wales, a CSIRO cattle research station some 30 km to the west of Potts Hill.

A key feature of this interferometer was that Bernie independently developed a phase switching technique essentially the same as that developed earlier by Martin Ryle, the leader of the radio astronomy team at Cambridge. One of the main issues he had been dealing with was that the relatively low sensitivity of the interferometer meant that it required long periods of integration and therefore the measurements suffered from gain fluctuation in the receiver. It was while seeking to overcome this problem that he recalled: 'I had begun to look for a better alternative when I received some unintended assistance from Cambridge. News came through the grapevine that a revolutionary system had been introduced there but it was all very hush-hush and no details were known; it was believed that it involved a modulation of the interference pattern. This seemed to be just what was needed. A little thought suggested modulation by interchanging maxima and minima on the interference pattern by switching phase and using a synchronous detector, as in the Dicke system. The necessary equipment was built and it worked very well. Later I found that this was precisely the system used at Cambridge, the only difference being their use of a hardware switch in the antenna feed lines whereas I had used an electronic switch following the preamplifier' (79).

Pawsey (1951) was careful to ensure that Mills acknowledged Ryle's priority in development of the phase switch, even though no technical details had been provided to the Radiophysics group. The interesting development in Bernie's phase switch was that by using an electronic phase reversal switch after the preamplifiers, it allowed a broader band implementation that would be necessary for the conceptual leap to the development of the cross array.

From February to December 1950 the team conducted a survey of the sky from +50° to -90° in declination using the new three-element interferometer (see Figure 2). They observed at 101 MHz on an east-west baseline with two spacings of 270 m and 60 m giving 40 arc min and 3 degree lobes, and identified 77 discrete sources. Although the initial purpose of simultaneously observing at two different aerial spacings was to remove any ambiguity as to which was the central lobe, an immediate indication could also be obtained of whether a source had any angular extent greater than 10 arc min in an east-west direction. This was evident by comparing the amplitude of the interference pattern on the larger spacing. Bernie noted that many of the sources were of an extended rather than point-source nature. He also noted that while the sources taken as a whole appeared to have an isotropic distribution, many of the stronger sources were concentrated along the Galactic plane and therefore actually fell into two major classes, Class I being Galactic and Class II being isotropic and either associated with 'radio stars' within the Milky Way or extra-galactic sources (9). In reaching his conclusions, Bernie also used the comparison of the source counts (log N) versus flux density (log S) to show that the Class II sources had a slope of -1.5, consistent with an isotropic distribution, while the Class I sources had a slope of -0.75, consistent with the non-homogeneous Galactic distribution. In conducting his analysis, he also pointed out some of the shortcomings of earlier attempts by Ryle (1950) and Bolton and Westfold (1951) to understand the distribution of radio emission. This marked the beginning of a long-running dispute with Ryle that would become known as the log ($N$) – log ($S$) controversy (79).

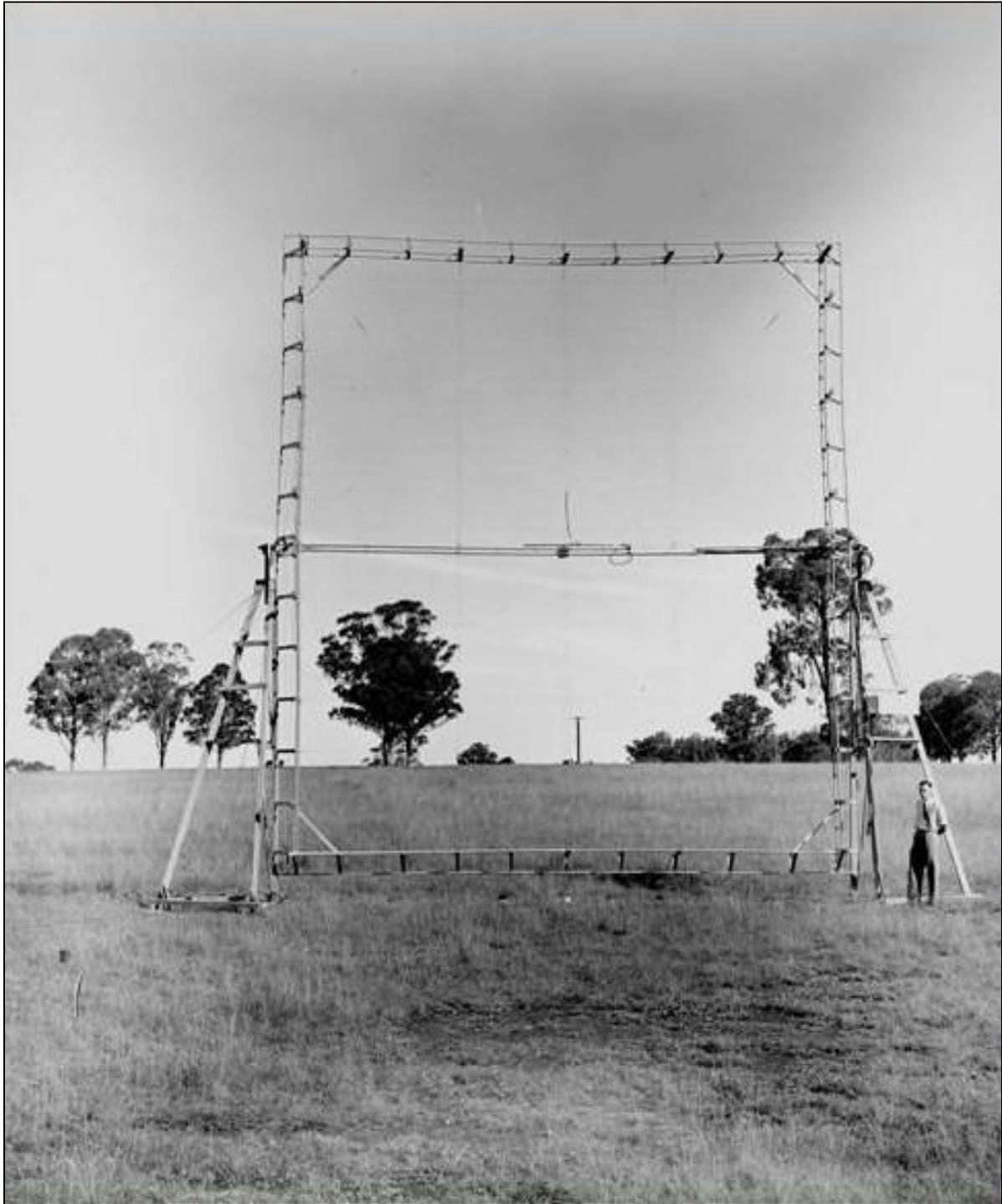
Figure 2: One of the three broadside antennas of the 101 MHz interferometer at the Badgerys Creek field station (Courtesy of CSIRO Radio Astronomy Image Archive B2095-1 Image Date: 9 May 1950).

Bernie next conducted a more detailed investigation of six of the stronger discrete sources to determine accurate position estimates (8) and to examine their angular extent on the E-W baseline. To obtain 1 arc min resolution at 100 MHz, he needed to extend the maximum baseline of his interferometer to 10 km. To achieve this he developed a radio link, transmitting both the signal and local oscillator frequencies to preserve the phase of the signal when reconstituting the signal at the receiving link. This was the first time this technique had been used in radio astronomy and it was quickly adopted by the Jodrell Bank group who were also exploring the angular size of discrete sources.

With the 1952 URSI (International Union of Radio Science) conference rapidly approaching, and with both the Jodrell Bank and Cambridge groups also studying the nature of the discrete sources, Mills decided to concentrate on resolving the sources Cygnus A, Taurus A, Virgo A and Centaurus A and on trying two other baseline orientations to help elucidate the angular extent of the sources. Completing these observations just in time for the URSI meeting, Bernie found that the angular radio sizes were comparable to the optical sizes of the nebulae themselves and this provided strong evidence that the discrete sources were largely, if not entirely, nebulae of both galactic and extra-Galactic origin. The results were presented together with results from Jodrell Bank and Cambridge during URSI 1952 and later published in *Nature* (10) (Hanbury-Brown et al., 1952; Smith, 1952). In the more detailed follow-up paper, Mills (13) produced 'radio pictures' for comparison with the optical nebulae counterparts. This was the first time a grey-scale representation of brightness distribution had been used in radio astronomy. Unfortunately, as Sullivan (1982, p. 274) has noted, in his rush to complete the observations he lacked the critical spacing in his observations between 1 and 5 km that would have allowed him to detect for the first time the double structure of Cygnus A, and that would later prove the key to understanding the nature of these sources. This honour would go to the Jodrell Bank team (Jennison and Das Gupta, 1953) who also used a radio-link interferometer but with a large sample of spacings that allowed them to detect the double-lobe nature of the source.

These first years of the 1950s marked the beginning of a revolutionary period in understanding the nature of the discrete sources and would set Bernie on a path of challenging assertions, particularly from Martin Ryle's group in Cambridge, on the nature of 'radio stars' and the distribution of the discrete sources and their cosmological implications.

**The Idea for a Cross**

Using the new phase-switched interferometer at Badgerys Creek, Bernie soon discovered another major problem in using spaced interferometers for survey work. Many of the sources resolved at the short spacing bore no resemblance to those detected at the longer spacing. He had determined that this was most likely caused by the sources being extended in nature rather than being point sources.

Bernie believed that sensitivity was not the issue for the source survey work at metre wavelengths; rather, high resolution was the key requirement. As he has stated: 'By then, I knew that collecting area was relatively unimportant; the important thing was a large overall size to give high resolution. As a filled array seemed wasteful, I first looked at various passive configurations such as crosses and rings, but these all suffered from high side-lobe levels. Suddenly it occurred to me that by combining the phase-switch, which I had used on the interferometer, with a crossed array the side-lobe problem would be substantially reduced.' (79)

The extent of Bernie's physical grasp of how things work is well illustrated by his commentary on the evolution of the cross concept: 'A solution occurred to me after discussing the imaging problem with Christiansen who was using two grating arrays along the sides of a reservoir to produce maps of the Sun by the first application of earth rotation synthesis. However, fast imaging was really needed because of the variable solar emission, quite apart from the inconvenience of carrying out Fourier transforms when no computer was available. With my thoughts concentrated on linear arrays I soon realized that a solution to both our needs was an antenna in the form of a symmetrical cross, with the outputs of the arms combined through a phase reversing switch as then used in my interferometer systems. Only the signals received in the overlapping area of the fan beams would produce a modulated signal that could be picked out with a phase-sensitive detector to produce a simple pencil beam response or, in the case of grating arrays, an array of pencil beams. This process effectively multiplied the two antenna responses'.

This conceptual breakthrough went back to Bernie's earlier experiences with the phase-switched interferometer with two elements set to slightly different declination angles, so that a primary beam null sat on an interfering source. The product of the signals from the two elements (produced by phase-switching one, adding and passing to a square-law detector and phase-sensitive detector) had a null on the interfering source but a strong signal from the wanted source. This multiplication of the beam responses was precisely the technique needed to produce the principal beam response of the cross.

While Pawsey quickly embraced the idea of constructing a cross, others were more sceptical and it was therefore determined that a proof of concept should be constructed. Bernie was joined by the Division's Technical Officer, Alec Little, at the start of what would turn out to be a 32-year partnership. They built the prototype at Potts Hill using chicken wire and wooden posts, with the arms of the cross 36 m in length (Figure 3). The combined response of the arms produced an 8° pencil beam. The prototype not only successfully proved that the concept was viable, it also resulted in the first radio-continuum detection of the Large Magellanic Cloud (14).

The cross concept proved versatile in radio astronomy and also later inspired new techniques in radar and, more widely, in multi-beam sonar designs. In this case, one line array was used for transmission and another at right angles (not necessarily forming a cross) for reception so that the combination provided a 'beam'. The first of these was patented in 1964 by SeaBeam Instruments—at the time the Harris Anti-Submarine Warfare Division of General Instrument Corporation—and was used by the US Navy and later by the Royal Australian Navy [L-3 Communications SeaBeam Instruments, 2000]).

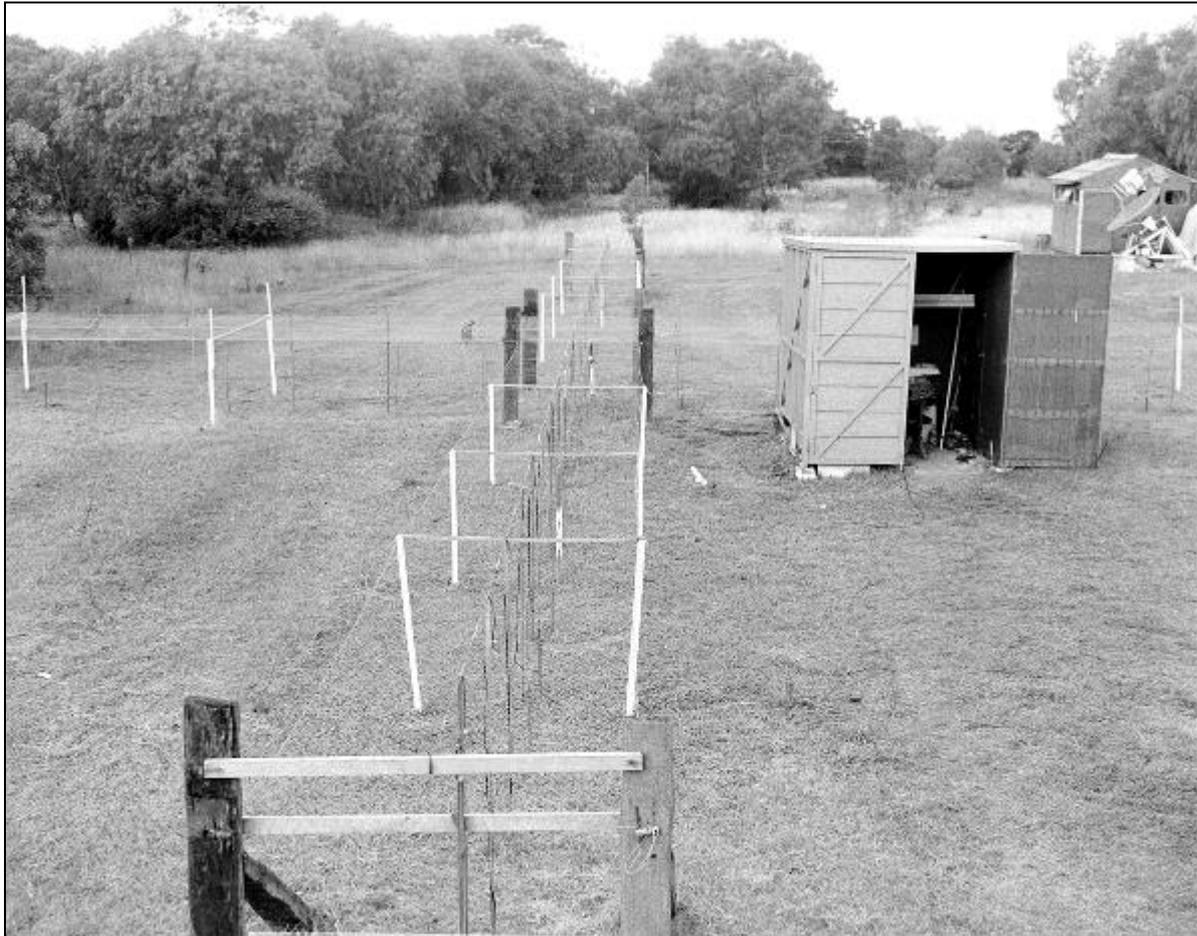

Figure 3: A close up view of the prototype Mills Cross at the Potts Hill field station. The prototype cross operated at 97 MHz with a beamwidth of 8 degrees. Each arm of the cross was 36m in length (Courtesy of CSIRO Radio Astronomy Image Archive: B3064-3 Image Date: 21 April 1953).

## Fleurs and the Mills Cross

Following the success of the prototype cross at Potts Hill, Radiophysics allocated the necessary resources to build a full-scale instrument. A new site was needed as both Potts Hill and Badgerys Creek had insufficient flat ground for such an instrument. A disused Second World War airstrip close to Badgerys Creek provided the ideal location and in 1953 the Fleurs field station was established.

The new cross was constructed during 1953-4, largely by Alec Little as Bernie was on a six-month study tour of the USA: 'During all this activity I received an invitation to spend six months in the United States visiting the California Institute of Technology and the Department of Terrestrial Magnetism [Carnegie Institute of Washington], which was developing a program in radio astronomy. The invitation came at an awkward time, but to decline was unthinkable and I had no qualms about leaving the supervision of construction in the capable hands of Alec Little. This visit was well worthwhile as the few months spent at Cal Tech marked a turning point in my grasp of astronomy and astrophysics. Discussions with some of the leading astronomers and astrophysicists of the day (particularly the iconoclastic Fritz Zwicky), attendance at colloquia, and even a postgraduate course on stellar structure all helped to fill in some of the numerous gaps in the knowledge I had managed to acquire. I returned home in early 1954 with my mind full of plans for observational programs' (92).

The new cross operated at 85.5 MHz, with each of the north-south and east-west arms 450 m in length, giving a beamwidth of 48 arc min (Figure 4). Observations began in 1954 and the cross was immediately successful, with studies of Galactic sources, the Magellanic Clouds and other external galaxies.

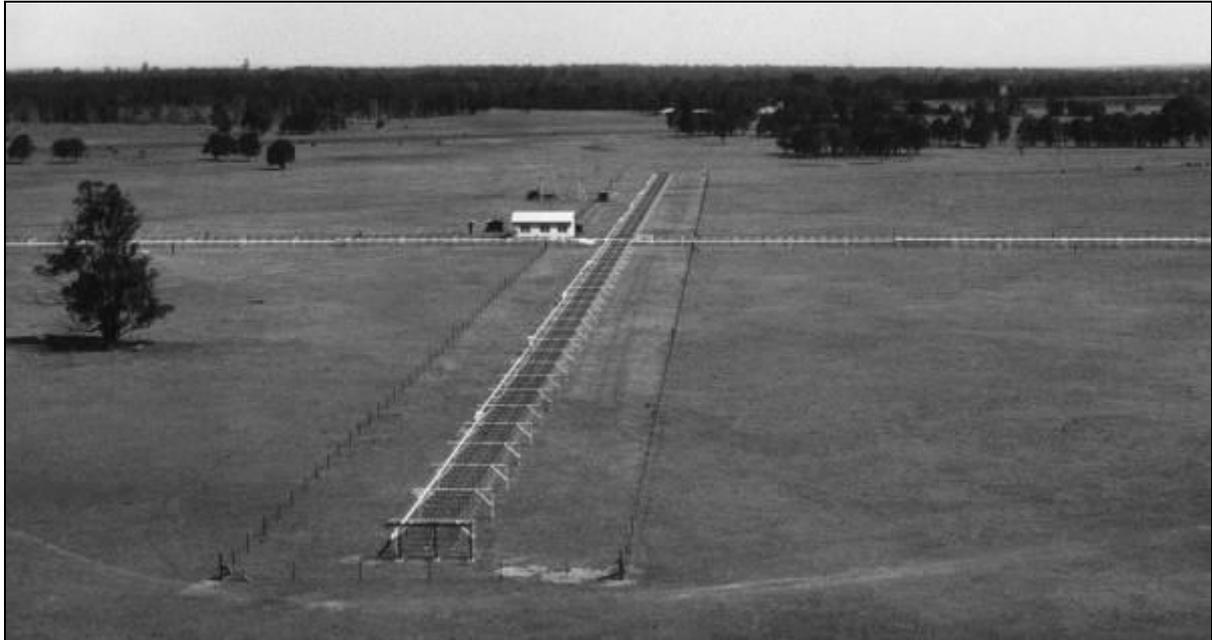

Figure 4: The Mills Cross at the Fleurs field station looking south along the N-S arm. The telescope operated at 85.5 MHz with a beamwidth of 48 arc min. Each arm of the cross was 450m in length (Courtesy of CSIRO Radio Astronomy Image Archive: 3476-3 Image Date: 25 October 1954).

**Controversy with Cambridge**

Soon after beginning a southern-sky survey with Bruce Slee using the new cross, Bernie received a letter from Fred Hoyle asking his views on the validity of the data and cosmological claims being made by Ryle's group based on their Second Cambridge (2C) survey (Shakeshaft et al., 1955). By this time Ryle had reversed his earlier views that 'radio stars' were the source of the discrete radiation and now favoured an extra-galactic origin. Based on the newly completed 2C survey, the Cambridge team found that examining the plot of the logarithm of the number of sources (*N*) versus the logarithm of flux density (*S*) produced a slope steeper than -1.5 based on a very large number of faint sources. A slope of -1.5 would have been consistent with a Euclidean universe and the Steady State cosmological theory proposed by Hoyle. In his 1955 Halley lecture at Oxford, Ryle had audaciously concluded that his team's data effectively ruled out the Steady State model (Ryle and Scheuer, 1955; Ryle, 1955). This pronouncement marked the entry of radio astronomy into the field of cosmology.

Bernie later recounted in detail the events that unfolded around the discord between the two surveys (79) and the circumstances have been discussed extensively by Edge and Mulkay (1976) and Kragh (1996). While initially there was regular communication with Cambridge about the discrepancies that were already apparent between the surveys, the relationship quickly deteriorated as Ryle refused to accept that there were serious problems with the 2C survey. After it became clear that Ryle was ignoring the conclusions being reached in Sydney, Bernie decided to publish their own preliminary survey results, including a formal criticism of the Cambridge survey, pointing out the flaws in technique and that 'deductions of cosmological interest derived from its analysis are without foundation' (21). With the benefit of hindsight it is clear that the 2C survey was severely affected by 'confusion' (the blending of several weak sources in the side-lobes of the interferometer to give a response like one stronger source) and that Bernie's objections to drawing conclusions based on the counts of discrete sources were correct. Largely lost in the debate on source counts, the analysis of the 2C probability distribution $P(D)$ of the interferometer deflection amplitudes ($D$) proposed by Peter Scheuer at the time proved to be a valid method and confirmed Ryle's conclusion that there was a real increase in the density of sources at low flux density, implying source evolution in the distant past. However, it also showed that the log ($N$)-log ($S$) slope would be much less steep than the -3 slope found by the Cambridge team based on the source counts (Scheuer, 1990). By late 1956, the observations for the Third Cambridge (3C) survey were largely complete. Early analyses of them confirmed the criticism of gross confusion in the 2C catalogue although the final results were not published until 1959 (Edge et al., 1959). Ryle knew by late 1957 from the preliminary 3C results that the log (N)-log (S) found in 2C was unreliable but he refused to withdraw his claims. The debate between cosmological models would continue for several more years until it was largely settled by the serendipitous discovery of the cosmic microwave background by Penzias and Wilson in 1965.

While the controversy continued, the Mills Cross at Fleurs proved itself to be an excellent survey instrument at metre wavelengths and especially suited for the study of our own galaxy, allowing the different components of galactic radiation to be identified. The resulting survey was however itself not immune from the effects of confusion, with some of the extended sources later proving to be blends of separate sources. The full survey, known as MSH, was completed in 1957 and published in three separate papers (26, 35, 37/38). The survey at 85.5 MHz, with a 48 arc min beam, covered most of the southern sky between declinations of +10° and -80°, cataloguing 2270 radio sources down to around 7 Jy. The MSH catalogue remains a largely untapped resource. Complete at such a low frequency, it could have been used to build a southern complement to the revised 3C survey (178 MHz) which has provided the unique sample of powerful steep-spectrum sources so heavily studied by northern-hemisphere astronomers and from which such fundamental knowledge of Active Galactic Nucleus (AGN) physics and cosmology flowed.[5]

In an attempt to understand the nature of the extragalactic discrete sources better, Bernie sought to derive a luminosity function (34) and to measure their angular extent more accurately (Goddard et al., 1960). Returning to the idea of the radio-link interferometer, the team extended the baseline by combining the cross with an aerial 10 km distant. In this configuration it was possible to obtain resolution as high as 10 arc sec at 85.5 MHz.

---

[5] The authors are grateful to Jasper Wall for his comments and discussion of the significance of the MSH Survey.

However, in both cases the findings were inconclusive and a more detailed understanding would have to wait for the discovery of quasars, which are barely detectable at 85 MHz due to their flat and inverted radio spectra. Higher-frequency surveys were needed to begin to understand the nature of these sources.

In 1959, Bernie was awarded the degree of Doctor of Science in Engineering by the University of Sydney for a thesis covering the development of the 85.5 MHz cross and its observations. Coincidentally, one of the authors of this paper (Bob Frater), an undergraduate student at the time, had stood in awe at the back of the University of Sydney Great Hall during the award ceremony and would later go on to be awarded the same degree for his work on the Molonglo Cross and the Fleurs Synthesis Telescope, very much a legacy of Bernie's engineering prowess.

With the MSH Survey now complete, Bernie travelled for the first time to Europe to attend the Paris Symposium on Radio Astronomy and then the 1958 International Astronomical Union General Assembly in Moscow where he presented many of the results from the observations covering Galactic structure and the discrete sources. This chance to meet and present his own research to the broader astronomical community marked the end of his successful transition from engineer to astronomer.

**Schism at Radiophysics**

Bernie's return from Europe marked not only his transition to a fully-fledged astronomer but the beginning of the end of his time with the CSIRO Division of Radiophysics. Despite the versatility and performance demonstrated by the cross-type radio telescopes, a decision was made to abandon the Division's fifteen-year tradition of astronomy using ingenious small aerial systems and, in particular, any further development at Fleurs. The era of large parabolic dishes had already begun in England with the 250ft (76m) Jodrell Bank Radio Telescope. At Radiophysics, planning of a 'Giant Radio Telescope' (GRT) had begun in 1953 (Robertson, 1992, p. 133). Bowen succeeded in obtaining funding from both the Carnegie Corporation and the Rockefeller Foundation in the USA for a GRT in Australia and the design contract for it was placed in 1956. When Mills and Christiansen drove the peg to mark its approximate position on a farm near Parkes (Figure 5), both knew that there was no future within Radiophysics for their types of cross and array radio telescopes in cosmic research.[6] All remaining resources available to the Radiophysics Division were reserved for Paul Wild's radioheliograph at Culgoora, New South Wales, for solar research and there was nothing left for a large cross (Bowen, 1981). Bernie's criticism of the GRT project was not about the scientific usefulness of the telescope, which proved extremely well suited to high frequencies and spectral analysis, rather it was that Radiophysics would effectively be abandoning high-resolution, low-frequency cosmic research (92).

---

[6] As described in letters from Bowen to Freeman Fox & Partners and to Pawsey on 20th March 1958, Bowen left the selection of the site of the GRT site to the "radio astronomers" with Mills and Christiansen marking the site in March 1958. As an interesting aside, Bowen would later recreate the event of driving in the first peg at the Parkes site at another location some distance to the south of the telescope (see Figure 5).

The planning and commitment to the GRT at Parkes triggered major career changes for Bernie and many other Radiophysics staff who left for research positions elsewhere. Bernie investigated chairs of Electrical Engineering at Adelaide, Melbourne and Sydney but these did not offer the financial support he needed to build his large, cross-type telescope. He found this support in 1960, not in Engineering but in Physics at the University of Sydney.

Also during this period of turbulence, Christiansen left Radiophysics for the University of Sydney while Pawsey, intending to become the second Director of the National Radio Astronomy Observatory in the USA, became ill and died in 1962. Pawsey had been a major proponent of a new Mills Cross with higher resolution and sensitivity. He called this proposed instrument 'the Super-Cross' and noted that Mills' past work 'has been outstanding, his contribution has probably been the greatest single factor in giving Australian radio astronomy the high prestige it now enjoys' (Pawsey, 1960). One of the reasons Pawsey left Radiophysics was his frustration at being unable to find funding for the new project.

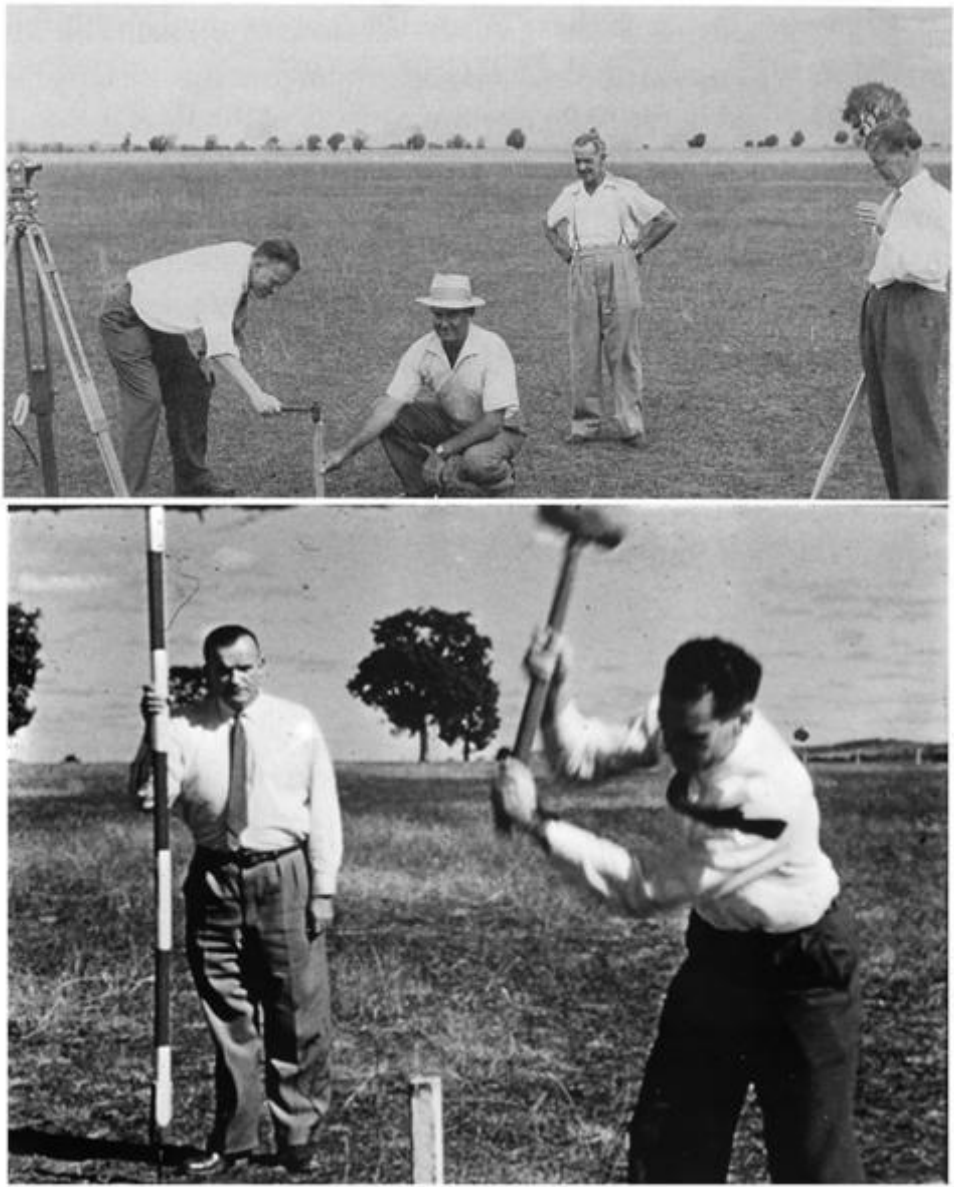

Figure 5: Top - The original record, taken in March 1958, of Mills hammering in a survey stake at the Parkes site with Christiansen, to the far right, looking on (courtesy of the Mills family archive). Bottom – The official CSIRO film documenting the Parkes Radio Telescope required a starting piece. This led to the recreation of the peg driving by Bowen that was witnessed by local farmer Phil Jelbart in September 1961 just prior to the official opening. The second image is believed to be a still from this film created in the Radiophysics photographic laboratory on 29 September 1961 by which time Mills and Christiansen had moved to Sydney University. It shows Bowen driving in the stake watched by Lindsey McCready (courtesy of CSIRO Radio Astronomy Image Archive B6586).

**Start of Molonglo Planning**

Harry Messel had been appointed as Professor of Physics at the University of Sydney in September 1952 and established the Nuclear Research Foundation (later the Science Foundation for Physics) to 'promote, foster, develop and assist ... research with grants from … fees, donations and the like' (McAdam, 2008). Between November 1959 and November 1961 Messel recruited new professors in theoretical physics, thermo-nuclear (plasma) physics and high-energy nuclear (that is, cosmic-ray) physics, as well as in electronic computing. When Robert Hanbury Brown in Manchester sought funds and a site for his optical intensity interferometer, Messel began an astronomy group. Richard Twiss, Cyril Hazard and John Davis, who had earlier been recruited by Messel, joined Hanbury Brown to build the intensity interferometer at a site near Narrabri in northern New South Wales. Messel also had funds for a complementary photometric telescope and sent Colin Gum to Europe to examine optical designs. Unfortunately, in April 1960 Gum died in a skiing accident in Switzerland and the telescope project never went ahead. Messel contacted Mills, approved the concept of a large cross-type radio telescope and offered him seed money sufficient to build a 408 MHz Cross with arms about 400 m long.

Thus in June 1960, at age 40, Bernie started a new career at the University of Sydney where he was to remain for the next 25 years. His initial plans were dependent on funding. He later commented: 'From the beginning there seemed to be few problems in constructing a Cross within the available budget of $200,000 which would be able to survey the sky at metre wavelengths with a sensitivity and resolution at least equal to that anticipated for the Parkes radio telescope operating at its optimum wavelength. But why stop there?' (89). Additional funds would mean longer arms replicating a flexible modular design. The challenge was to find the additional financial support for a larger cross.

Through some of his many overseas contacts (probably Thomas Gold at Cornell University), Messel learned that the National Science Foundation (NSF) was willing to make grants outside the USA. Mills quickly wrote a proposal for his ambitious 1-mile cross-type radio telescope. In support of this he provided results from his 85.5 MHz Fleurs Mills Cross survey and made precise predictions of possible observing programmes, the number of fainter sources expected, their confusion levels and the sensitivity required of the telescope.

The project met opposition from Bowen who advised against any grant, stating that a small university group could never manage such a large project.[7] When the NSF sent Geoffrey

---
[7] Letter from Mills to G. Keller, 13 December 1961; University of Sydney Archives, G047.

Keller, their Project Director for Astronomy, to Australia to investigate, Messel advised him to go to Canberra and talk to Bart Bok, the director of Mount Stromlo Obervatory. The visit reassured the NSF and in 1962 they approved the grant. The initial funding of US$746,000 was followed by a further US$107,500 in 1964 and allowed the Molonglo Cross project to go ahead with its planned mile-long arms. In his first published description of the project, Mills (39) wrote: 'This is a greatly enlarged version of the original "Mills Cross" put into operation by the CSIRO in 1954 ... the beamwidth would be about 2.8 arcmin and the sensitivity adequate to detect more than a million radio sources.'

Meanwhile Messel negotiated the purchase of a site for the new Cross in a wide flat valley near Canberra. This was one of the sites that had been investigated for the GRT but that had been rejected in favour of Parkes. The height of the GRT would have put it in line of sight over hills to the radio transmitters on Black Mountain in Canberra, but the cylindrical reflectors of the Cross were lower and remained shielded. Thus in the Parish of Molonglo on a branch of the Molonglo River, construction of the Molonglo Radio Observatory was commenced in 1961 (Figure 6).

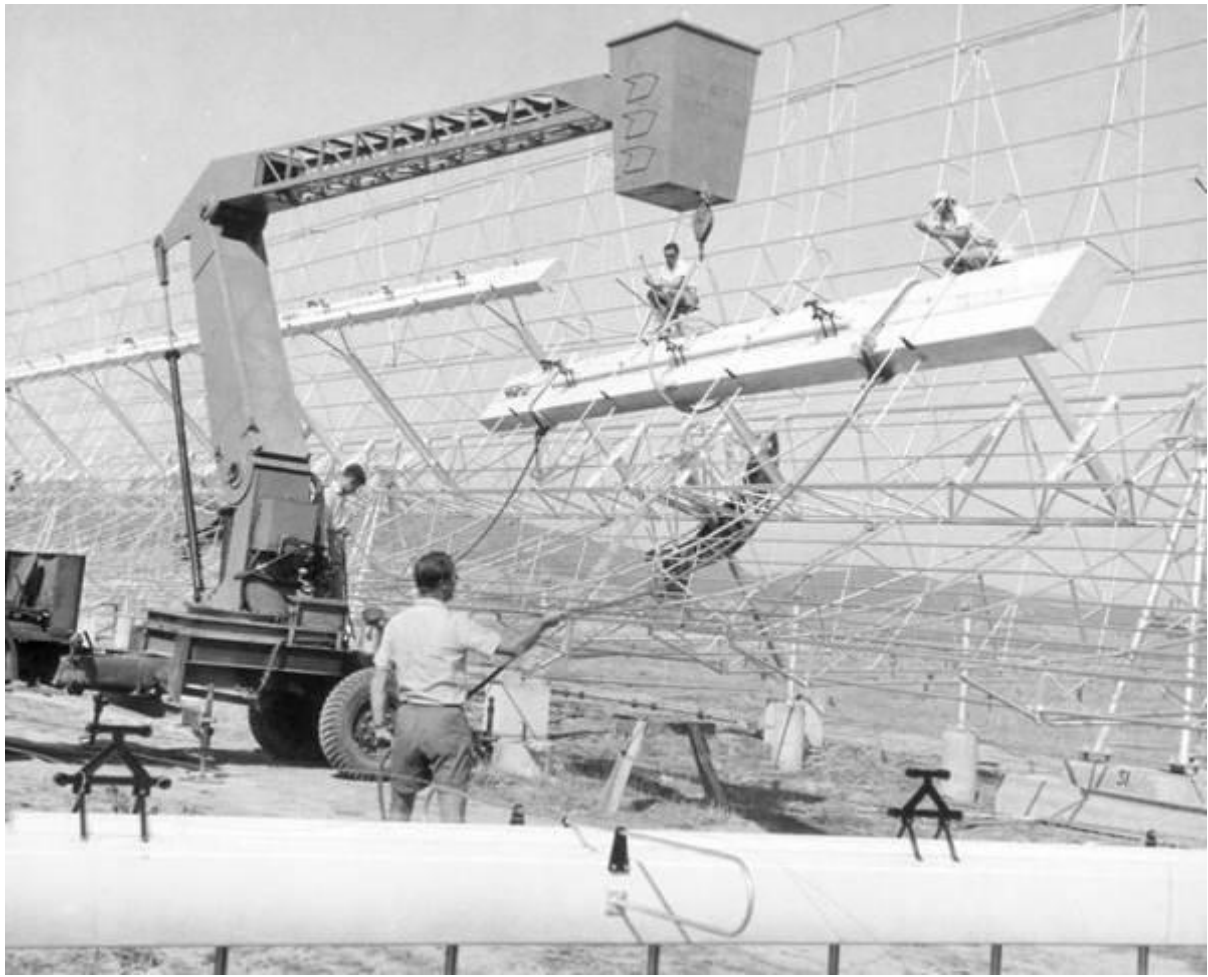

Figure 6: The construction of the east-west arm of the Molonglo Cross (courtesy of the University of Sydney Archives).

Bernie brought two valued colleagues from Radiophysics, Alec Little and Arthur Watkinson, and recruited three young radio astronomers from the UK, Bruce McAdam and Tony Turtle from Cambridge and Michael Large from Manchester. This small university group built the Cross over the next six years, but did so in co-operation with many university and industrial colleagues. Many years later, Mills (92) was to reminisce: 'I found myself manager of a big engineering project. It was not an enjoyable job but there was no one else to do it and I was much helped by my engineering contacts, stretching back in some cases to student days.'

From the start there was a major partnership with Christiansen in the School of Electrical Engineering at Sydney, who took responsibility for the receiving system. The co-operation was made formal with the formation of the Radio Astronomy Centre in the University of Sydney (Messel, 1960). One of us (Bob Frater) was enticed to leave industry and join the Electrical Engineering Department in 1961 specifically to work on the electronic design of the Molonglo Cross using the (then new) transistor technology. He later (2005) commented: 'I jumped at the opportunity. Bernie had in mind an instrument where the technical demands stretched significantly beyond the technology of the time.'

The official opening ceremony for the Molonglo Observatory was held on 19 November 1965 and was attended by the then Prime Minister of Australia, Sir Robert Menzies (see Figure 7).

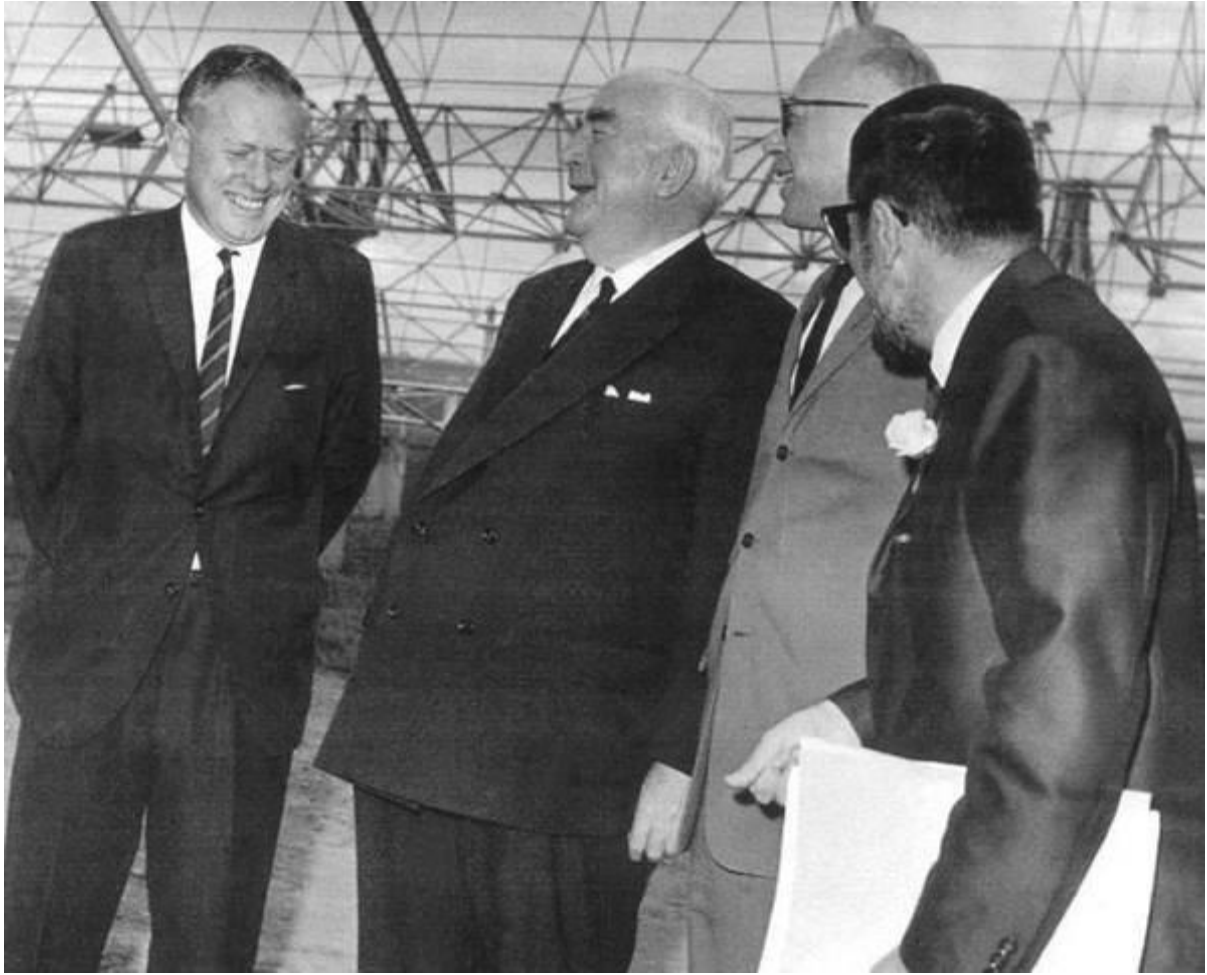

Figure 7: Mills enjoying a moment with the then Australian Prime Minister, Sir Robert Menzies during the official opening ceremony for the Molonglo Observatory on 19 November 1965. The Head of the School of Physics at the University of Sydney, Professor Harry Messel, is on the far right (courtesy of the University of Sydney Archives).

The full cross became operational in 1967, making measurements of the sky at 408 MHz (Figure 8). Major research achievements using the new telescope during its eleven years in operation included the Molonglo Reference Catalogue of 12,000 sources and numerous pulsar surveys. More than half the pulsars and supernova remnants known at the time were discovered at Molonglo, the most significant (see Figure 9) being the Vela pulsar . The accurate position provided enabled the subsequent association of this with the Vela supernova remnant and the later optical detection of the pulsar using the Anglo Australian Telescope (Wallace et al., 1977; Goss et al., 1977). The accurate Molonglo positions (typically 2-3 arc sec) enabled radio-optical identifications to be established on the basis of positional agreement alone, thereby providing a grid of southern-hemisphere calibrators for both Molonglo and Parkes. The Molonglo Cross made a complete survey of the southern sky, including the Galactic Plane, that showed diffuse radio emission delineating the spiral arms of the Galaxy as first seen in Bernie's early MSH survey, and an iconic image of the Galactic Centre.

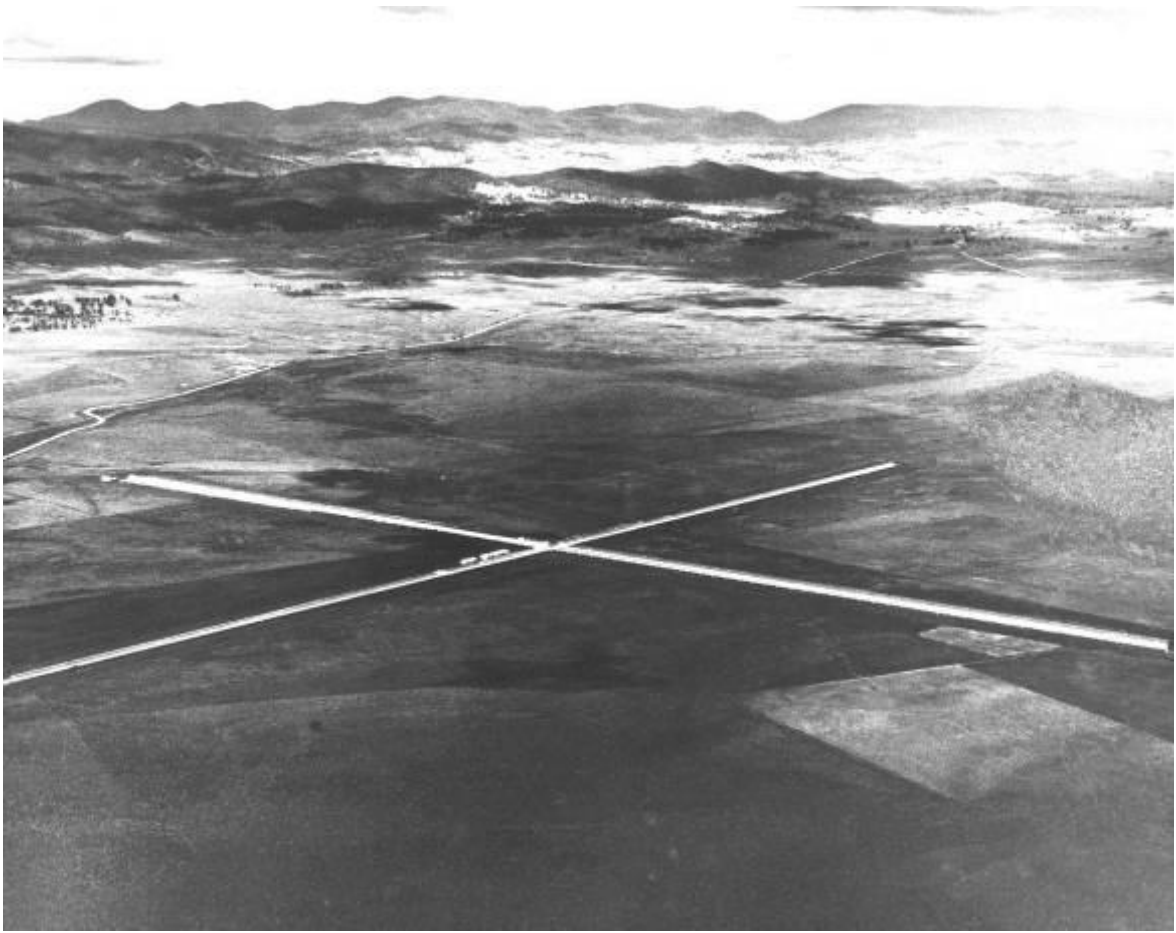

Figure 8: An aerial view of the completed Molonglo Cross looking south-east. The observatory buildings visible near the centre of the cross are aligned along the north-south arm. The telescope operated at 408 MHz with a beamwidth of 2.8 arc min. Each arm of the cross was 1.6km in length. (Courtesy of the University of Sydney Archives).

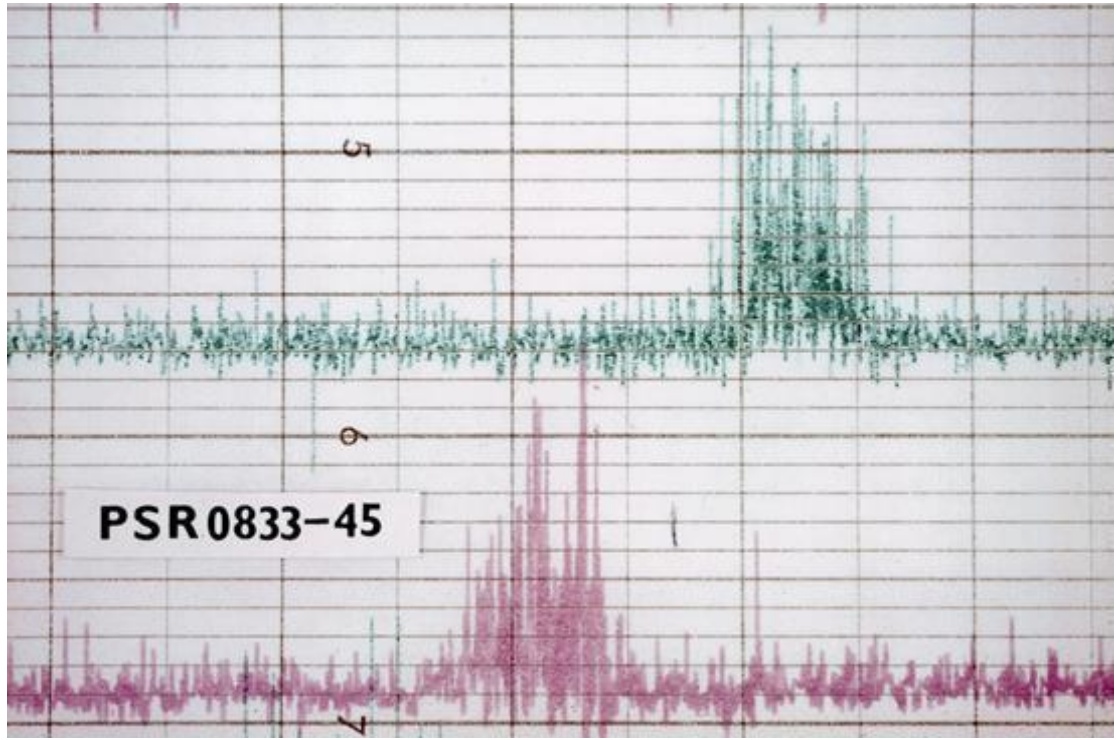

Figure 9: A chart record of the short-period Vela pulsar discovered with the Molonglo Cross operating at 408 MHz with a beamwidth of 2.8 arc min in 1968 (courtesy of the University of Sydney Archives).

During the development of the Molonglo Observatory, Bernie managed to attract an amazing array of talent to the project both directly and indirectly. This was testament not only to his reputation as an astronomer and engineer but also to his skill as a leader. Among the University of Sydney staff associated with Molonglo Observatory, 1960-78, were:

*From the School of Physics:* David Crawford, John Durdin, Richard Hunstead, Michael Large, Alec Little, Alan Le Marne, Hugh Murdoch, Bruce McAdam, Tony Turtle and Arthur Watkinson, with technical support from Terry Butcher, Grant Calhoun, John Horne, Jack Howes, Cornelius Kohlbrugge and Michael White.

*From the School of Electrical Engineering:* Ron Aitchison, Chris Christiansen, Bob Frater, Ian Lockhart and Cyril Murray.

Conversion of the Molonglo Cross to the Molonglo Observatory Synthesis Telescope began in 1978 and operations began only three years later. Bernie has emphasized the essential role of Alec Little in this conversion (85); Alec died in March 1985, aged 60, before the full potential of his work could be achieved. Bernie lost a valued colleague who had worked with him for thirty years.

**The Molonglo Observatory Synthesis Telescope (MOST)**

By the early 1970s, digital computers had achieved both the speed and the reliability to take real-time control of a radio telescope. Bernie realised that if a fan beam tracked a field for twelve hours, the rotation of the Earth would move the beam through 180° on the sky and allow the synthesis of a pencil beam (65). He designed a synthesis telescope for 1,420 MHz and Alec Little had developed a prototype feed for the east-west modules, when they learnt that CSIRO was planning the Australian Synthesis Telescope (later the Australia Telescope) for this and higher frequencies. Bernie then chose a new frequency of 843 MHz which is not a protected radio astronomy band but, with co-operation from the Australian Post Office, has been kept reasonably free of interference from nearby fixed and mobile radio phone transmitters.

The conversion of the Cross to the MOST re-used the east and west arms and produced a powerful new telescope operating at 843 MHz that retained much of the original Cross infrastructure and electronics. This was the predominant reason for choosing the new operating frequency as it enabled reuse of the waveguides and local oscillator system. Mechanically, a slow tilt drive was added. The total length of the east and west arms together remained at 1.6 km (1 mile). The 408 MHz dipoles were replaced with 7,744 ring elements that were phased by differential rotation under computer control to track a field for twelve hours. Black flower pots were a novel way of protecting the ring antennas (see Figure 10). The conversion of the feeds and construction of new receivers, digital delays and analogue beam formers to produce 128 contiguous fan beams took three years to complete (Robertson, 1991).

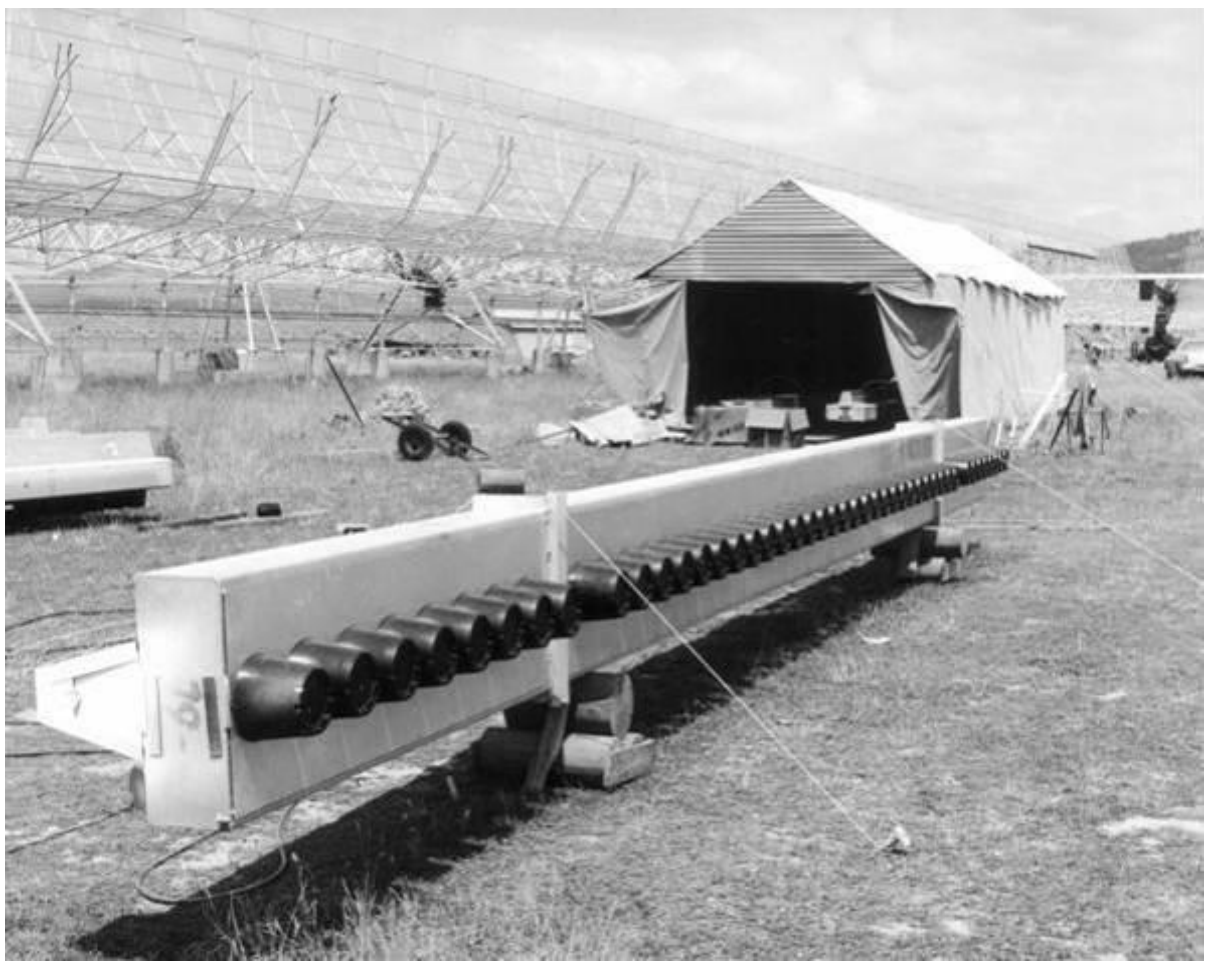

Figure 10: The construction of MOST showing the new feed elements designed to operate at 843 MHz. In the MOST configuration only the east-west arm (visible in the background) of the original cross was used (courtesy of the University of Sydney Archives).

The first synthesis map of source 1733-565 was made on 15 July 1981 with 43 arc sec resolution over a 23 arc min field. Switching beams across three adjacent centres increased the field to 70 arc min and detailed images of known sources up to one degree in size were observed for a decade.

The last major observation programme overseen by Bernie at Molonglo was a study of supernova remnants in the Magellanic Clouds (81). Bernie's career as an active astronomer and engineer ended in 1985 when he reached the then mandatory retirement age of 65. Bernie maintained an active interest in astronomy and radio telescope developments. Over time his interests shifted from the large-scale structure of the universe to the very small scale of the quantum world and an active interest in relativity theory (92).

In 1991, development of MOST continued with precise digital phase units (Amy and Large, 1992) giving computer control of phase for all 176 modules, thus removing grating lobes and giving a great improvement in dynamic range over the 70 arc min field. A further installation of phase control to the separate waveguides within each module in 1996 increased MOST's field of view to more than 5 square degrees. With the large field of view, it became feasible to undertake an 843 MHz survey of the southern sky from declination -30° to -90°. The Sydney University Molonglo Sky Survey (SUMSS) began in 1997 and finished in August 2007 (Bock et al., 1999; Mauch et al., 2003). A parallel project (a second-epoch Molonglo Galactic Plane Survey; MGPS-2) mapped the southern sweep of the Galactic Plane through to the Galactic Centre (Murphy et al., 2007). Both surveys are available online in catalogue format and as images. Through the 1980s and early 1990s, MOST was used for targeted observations of extended radio galaxies, clusters of galaxies, the Magellanic Clouds and supernova remnants. Discovery of the prompt radio emission (see Figure 11) from supernova 1987A (Turtle et al., 1987) and its subsequent long-term monitoring remains a highlight, along with the rapid-response detections of radio emission from transient Galactic X-ray sources.

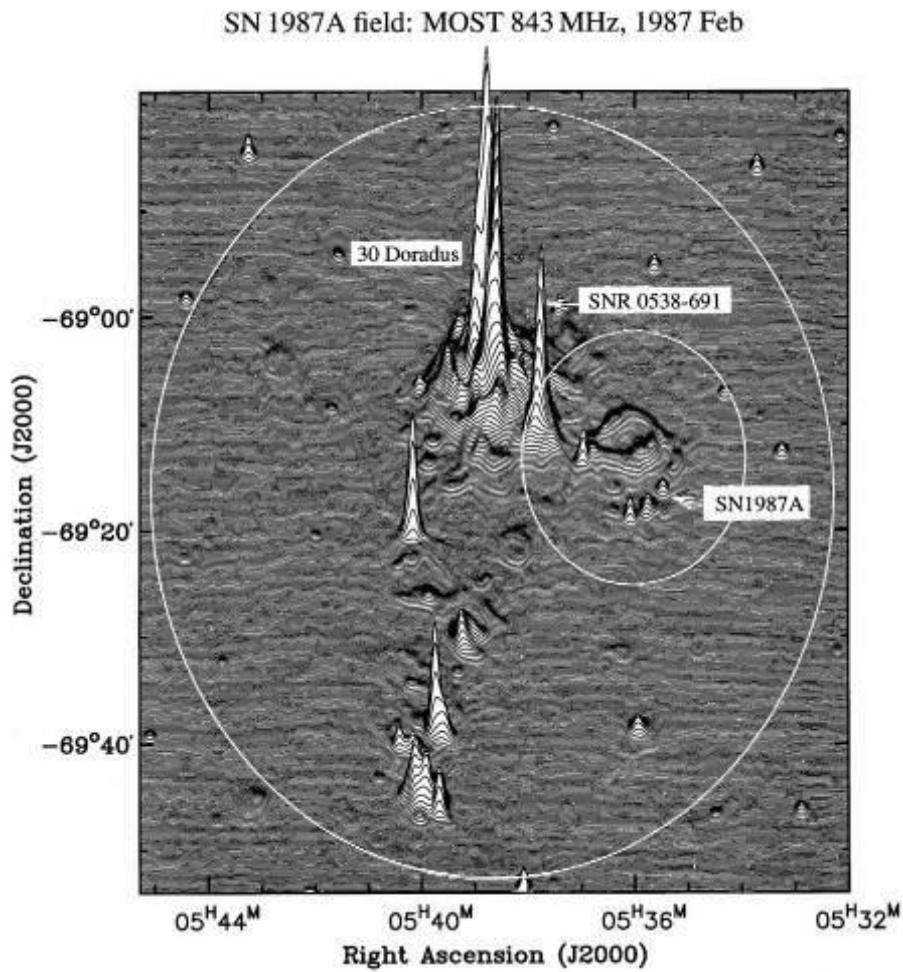

Figure 11: The MOST discovery image of prompt radio emission from Supernova 1987A at 843 MHz with a beamwidth of 45 arc sec and a 70 arc min field of view. The peak flux density at around 1 GHz was about 150 mJy and occurred within four days of the Supernova (after Turtle et al., 1987).

Following completion of SUMSS and MGPS-2, the telescope was funded for a transformational upgrade to trial a spectroscopic capability and to test the feasibility of polarization measurements using cylindrical antennas. New technology developments include wideband feeds, digital electronics (receivers, filterbanks and correlator) based on programmable logic chips and fast signal processing. The project was approved as one of the pathfinder instruments for the next-generation radio telescope, the Square Kilometre Array (SKA), and is called the SKA Molonglo Prototype (SKAMP), a joint CSIRO / University of Sydney venture. The existing mechanical infrastructure of the Molonglo telescope has again been used, giving a renewed life well beyond Bernie's original estimates of fifteen to twenty years. The new signal pathway is testing digital signal processing while preserving the ability to produce high-fidelity images of faint, extended radio sources. Significant new investigatons may be possible if funding is found to complete SKAMP.

**The Person**

Bernie and Lerida had three children, Eric, Miranda and Deborah (later Shamynka). Lerida died in 1969.

Bernie was an intensely private and self-contained person, captured well in this comment from the eulogy delivered at his memorial service by his second wife Crys, née Lewis, whom he married in 1970: 'Bernie had two great passions in life. His research, his ever-questioning, ever-challenging mind was always open to new ideas. He also cared deeply for his students (though no doubt he didn't show it!) and he was thrilled when they succeeded. In our frivolous family conversations, he was usually silent, gently smiling. Then he would make one short, sharp, ironic comment which made us all laugh. He had a warm, amused smile and his laughter, when it came, was deep and soft. He was a modest, gentle, kind and unassuming man. Even as pain wracked him before death, he was saying over and over to us: "I'm sorry to be causing you all this trouble." And to me: "I'm sorry to be leaving you like this. I won't be able to look after you."'

For those who worked with him, Bernie was a great leader, and for those who related to his striving for physical understanding, a great mentor. His legacy lives on throughout the world in the lives of the many students and colleagues whose research he encouraged and guided. His absolute integrity as a scientist and friend remains an example that all of us can strive to emulate.

Professor Harry Messel, who recruited Bernie to Sydney's School of Physics, said at Bernie's memorial service: 'Bernie was the mentor of many well-known researchers, several of whom are with us today, a quiet but wonderful leader'. Many distinguished astronomers benefited from Bernie's astute mind and towering intellect. However, the legacy of Bernie Mills also lives on in his innovative telescope design that enabled important astronomical discoveries for a period far longer than the most optimistic expectations. These achievements are a tribute to his insight and vision. He will long be remembered for his immense contribution to low-frequency radio astronomy in Australia, and for the innovative design of his cross-type array which has found a wide application within and beyond the realm of radio astronomy.

**Honours and Awards:**

    1936   Dux of King's School, Sydney
    1941   BSc (University of Sydney)
    1943   BE (Second Class Honours) (University of Sydney)
    1950   ME (First Class Honours) (University of Sydney)
    1956   Fellow, Royal Astronomical Society
    1957   Lyle Medal, Australian Academy of Science
    1959   DSc (University of Sydney)
    1959   Fellow, Australian Academy of Science
    1963   Fellow, Royal Society of London
    1967   Britannica Australia Award for Science
    1976   Companion in the Order of Australia
    2003   Centenary Medal
    2006   Grote Reber Medal for Radio Astronomy

**Acknowledgements**

This memoir is based on the recollections of two of the authors' first-hand interactions with Bernie from the early 1960s until his death, Bernie's own unpublished biographical notes held in the University of Sydney Archives, the transcript of an interview with Bernie conducted by David Ellyard in 1977 that was kindly made available to us by David Ellyard, and discussions with many other people who worked with Bernie over the years. Ron Ekers, Anne Green, Bob Hewitt, Dick Hunstead, Ken Kellermann, Bruce McAdam and Jasper Wall all provided valuable comments on and input to earlier drafts of the manuscript. We are also grateful to three anonymous referees who provided comments, to the CSIRO for images provided courtesy of the CSIRO Radio Astronomy Image Archive, and to the University of Sydney Archives. The portrait photograph is published courtesy of University of Sydney Archives, and shows Bernie Mills in 1960 holding a sketch plan of the proposed 'Super Cross'. One of us (Goss) is supported by the National Radio Astronomy Observatory (USA) which is operated by Associated Universities, Inc. under co-operative agreement with the National Science Foundation.